\documentclass[aps, showpacs,showkeys,pra] {revtex4}
\usepackage{hyperref}
\usepackage{amsmath }
\usepackage{amssymb}
\usepackage{epsfig}
\usepackage{natbib}
\newcommand*{\be}{\begin{equation}}
\newcommand*{\ee}{\end{equation}}

\begin{document}
\bibliographystyle{revtex}
\title{Rashba spin-orbit interaction in a circular quantum ring    \\
 in the presence of a magnetic field }
\author{V.V. Kudryashov}
 \email{kudryash@dragon.bas-net.by}
\author{A.V. Baran}
 \email{a.baran@dragon.bas-net.by}
\affiliation{Institute of Physics, National Academy of Sciences of
Belarus \\68 Nezavisimosti  Ave., 220072, Minsk,  Belarus }

\begin{abstract}{Exact wave functions and energy levels
are obtained for an electron in a two-dimensional semiconductor
circular quantum ring with a confining potential of finite depth
in the presence of both an external magnetic field and the Rashba
spin-orbit interaction.}
\end{abstract}

\pacs{03.65.Ge, 71.70.Ej, 73.21.-b}
\keywords{quantum ring, Rashba
spin-orbit interaction, magnetic field, exact wave functions and
energy levels}
\maketitle

\section{Introduction}

Quantum rings (nanorings) in  semiconductor heterostructures have
been an object of many investigations in recent years. Circular
quantum rings can be described as effectively two-dimensional
systems in a confining potential $V_c(\rho)$  ($\rho =\sqrt{x^2
+y^2}$). In contrast to the parabolic and infinite hard wall
models, in papers \cite{ban,gro}, a realistic model was proposed
in which an axially symmetric rectangular potential well
\begin{equation}
V_c(\rho) =  \left\{
\begin{array}{cl}
 V, & 0<\rho < \rho_i ,\\
 0, &  \rho_i < \rho <  \rho_o ,\\
 V, &  \rho_o < \rho < \infty
 \end{array} \right.
 \end{equation}
of finite depth $V = constant$ corresponds to a quantum ring,
where $\rho_i$ and $\rho_o$ are the inner and outer radii of the
ring, respectively.
 In \cite{ban,gro}, this potential  was applied in the presence of an
external magnetic field, however without taking into account the
spin-orbit interaction. Notice, that confining potential of finite
depth was used for the description of  quantum dots taking into
account the spin-orbit interaction in \cite{chap,kud1}.

 Spin-orbit
coupling in semiconductor nanostructures have attracted
considerable attention. The most studied spin-orbit interaction in
semiconductor structures is the Rashba interaction \cite{ras,byc}.
The vector potential ${\bf A} =\frac{B}{2}(-y,x,0)$ of an external
uniform constant magnetic field oriented perpendicular to the
plane of
 the quantum ring leads to the generalized momentum ${\bf P} ={\bf p} +
 q_e{\bf A}$,
where $q_e$ is the electron charge. The Rashba interaction is
described by the formula
\begin{equation}
V_R= a_R (\sigma_x P_y - \sigma_y P_x)/\hbar
\end{equation}
with standard Pauli spin-matrices $\sigma_x$ and $\sigma_y$. The
Rashba interaction can be strong in semiconductor heterostructures
and its strength can be controlled by an external electric field.

In \cite{kud2}, the potential (1)  was applied in the case of a
circular quantum ring in the presence of the Rashba spin-orbit
interaction, however without an external magnetic field. In the
present work, the quantum-mechanical problem  is exactly solved
for a nanoring in the  presence of both  the Rashba spin-orbit
interaction and an external magnetic field.

\section{Exact wave functions and energy levels}

The Schr\"odinger equation describing an electron in a
two-dimensional quantum ring normal to  $z$ axis is of the form
\begin{equation}
 \left(\frac{{\bf P}^2}{2 M_{eff}} + V_{c}(\rho) +V_R + V_Z \right) \Psi = E \Psi ,
\end{equation}
where $M_{eff}$ is the effective electron mass. We have the usual
expression for the Zeeman interaction
\begin{equation}
V_Z =\frac{1}{2}g \mu_B B\sigma_z ,
\end{equation}
where $g$ represents the effective gyromagnetic factor, $\mu_B$ is
the Bohr's magneton.

 The Schr\"odinger equation (3)
 is considered  in the cylindrical coordinates $\rho, \varphi$
 ($ x = \rho \cos \varphi,  y = \rho \sin \varphi )
 $.
 Further it
is convenient to employ dimensionless  quantities
  \begin{equation}
r= \frac{\rho}{ \rho_o}, \quad e =\frac{2 M_{eff}
\rho_o^2}{\hbar^2}  E, \quad
  v= \frac{2 M_{eff} \rho_o^2}{\hbar^2} V , \quad
  a=\frac{2 M_{eff} \rho_o}{\hbar^2}  a_R , \quad b = \frac{q_e  \rho_o^2}{2  \hbar} B ,
   \quad s = \frac{g M_{eff}}{4 M_e} .
 \end{equation}
Here $M_e$ is the electron mass. As it was shown in \cite{bul},
Eq. (3) permits the separation of variables
\begin{equation}
 \Psi_m(r,\varphi) =  u(r) e^{i m \varphi}\left(\begin{array}{c}
 1 \\
  0
 \end{array}\right) + w(r)  e^{i (m+1) \varphi} \left(\begin{array}{c}
 0 \\
  1
 \end{array}\right) ,\quad m=0, \pm 1, \pm 2, \ldots ,
\end{equation}
due to conservation of the total angular momentum $L_z +
\frac{\hbar}{2} \sigma_z $. An electron state is a linear
superposition of the states with the angular momentum numbers $m$
and $m+1$ which correspond to the opposite directions of spin.

We have the following radial equations
\begin{eqnarray}
 \frac{d^2u}{dr^2} + \frac{1}{r} \frac{d u}{d r}  +\left( e -v_c(r) - \frac{m^2}{r^2}
 -2 b m  - b^2 r^2   - 4 s b \right) u
= a  \left(\frac{d w}{d r} +  \frac{m+1}{r} w  + b r w\right) ,
\nonumber \\
  \frac{d^2w}{dr^2} + \frac{1}{r} \frac{d w}{d r}  +\left(e -v_c(r) - \frac{(m+1)^2}{r^2}
   -2 b (m+1)  - b^2 r^2  + 4 s b \right) w
= a  \left(-\frac{d u}{d r} +  \frac{m}{r}u  + b r u\right) ,
\end{eqnarray}
where
\begin{equation}
v_c(r) =  \left\{
\begin{array}{cl}
 v, & 0<r < r_i ,\\
 0, &  r_i < r <  1 ,\\
 v, &  1 < r < \infty,
 \end{array} \right.
 \end{equation}
 $r_i = \rho_i/\rho_o$.  In our model, we look for the radial wave functions
$u(r)$ and $w(r)$ regular at  the origin $r=0$ and decreasing at
infinity $r \rightarrow \infty$.

Following \cite{tsi} we use the substitutions
\begin{equation}
u(r) = \exp\left(\frac{-b r^2}{2}\right) (\sqrt{b} r)^{|m|}  f(
r), \quad w(r) = \exp\left(\frac{-b r^2}{2}\right) (\sqrt{b}
r)^{|m+1|} g( r) ,
\end{equation}
which lead to the confluent hypergeometric equations in the case
$a=0$. Therefore we attempt to express the desired solutions  of
Eq. (7) via the confluent hypergeometric functions when $a \neq
0$.

 We consider three regions  $0 <r < r_i $ (region 1),
$r_i <r <1$ (region 2) and $ 1 <r < \infty$ (region 3) separately.

In  the region 1, using the known properties
 \begin{eqnarray}
 M(\alpha,\beta,\xi) - \frac{d M(\alpha,\beta,\xi)}{d \xi} = \frac{\beta -\alpha}{\beta} M(\alpha,\beta + 1,\xi) ,
 \nonumber \\
 (\beta - 1  -\xi)M(\alpha,\beta,\xi) +\xi \frac{d M(\alpha,\beta,\xi)}{d \xi} = (\beta - 1)M(\alpha -1,\beta -1,\xi)
 \end{eqnarray}
of the confluent hypergeometric functions $ M(\alpha,\beta,\xi)$
of the first kind  \cite{abr} it is easily to show that
 the suitable particular solutions of the radial equations  are expressed via the functions
 \begin{equation}
f_1(r) = c_1^+ f_1^+(r) + c_1^- f_1^-(r) , \quad
 g_1(r) =  \left(\frac{a }{2 \sqrt{b}}\right) (c_1^+ g_1^+(r) + c_1^- g_1^-(r)) ,
\end{equation}
where
\begin{equation}
f_1^{\pm}(r) =  M(m+1-k^{\pm}_o,m+1,b r^2)  , \quad
 g_1^{\pm}(r)=
 \frac{k^{\pm}_o }{(m + 1)}
 \frac{  M(m+1- k^{\pm}_o,m+2,b r^2)} {(-k^{\pm}_o + (4
b)^{-1} (e - v) +s - 1/2)}
 \end{equation}
for $m=0,1,2...$ and
\begin{equation}
f_1^{\pm}(r) =  M(1-k^{\pm}_o,-m+1,b r^2)  , \quad g_1^{\pm}(r)= m
\frac{M(- k^{\pm}_o,-m,b r^2)} {(-k^{\pm}_o + (4 b)^{-1} (e -v) +s
- 1/2)}
\end{equation}
for $m=-1,-2,-3...$. Here the following notation
\begin{equation}
k^{\pm}_o = \frac{1}{4 b} \left( e - v + \frac{a^2}{2} \pm
 a \sqrt{e -v  +\frac{a^2}{4}
 + \left(\frac{4 b}{a}\right)^2 (s - 1/2)^2} \right)
\end{equation}
is introduced.  The corresponding wave functions have the
desirable behavior at the origin.

 In  the region 3, using the known properties
\begin{eqnarray}
U(\alpha,\beta,\xi) - \frac{d U(\alpha,\beta,\xi)}{d \xi} =
U(\alpha,\beta + 1,\xi) ,
\nonumber \\
(\beta - 1 - \xi)U(\alpha,\beta,\xi) + \xi \frac{d
U(\alpha,\beta,\xi)}{d \xi} = - U(\alpha -1,\beta - 1,\xi)
\end{eqnarray}
of the confluent hypergeometric functions $ U(\alpha,\beta,\xi)$
of the second kind  \cite{abr} it is simply to get the suitable
solutions of the radial equations   expressed through the
functions
 \begin{equation}
f_3(r) = c_3^+ f_3^+(r) + c_3^- f_3^-(r) , \quad
 g_3(r) =  \left(\frac{a }{2 \sqrt{b}}\right) (c_3^+ g_3^+(r) + c_3^- g_3^-(r)) ,
\end{equation}
where
\begin{equation}
f_3^{\pm}(r) =  U(m+1-k_o^{\pm},m+1,b r^2)  , \quad
 g_3^{\pm}(r)=
\frac{U(m +1- k_o^{\pm},m+2,b r^2)}{(-k_o^{\pm} + (4 b)^{-1} (e
-v) +s - 1/2)}
\end{equation}
for $m=0,1,2...$ and
 \begin{equation}
  f_3^{\pm}(r) =
U(1-k_o^{\pm},-m+1,b r^2)  , \quad
 g_3^{\pm}(r)=
 \frac{U(- k_o^{\pm},-m,b r^2)}{(-k_o^{\pm} +
(4 b)^{-1} (e -v) +s - 1/2)}
\end{equation}
for $m=-1,-2,-3...$ The corresponding wave functions  have the
appropriate behavior at infinity.

In the region 2, there are no restrictions on selection of the
particular solutions and the desired functions are determined as
follows
\begin{eqnarray}
 f_2(r) = c_{21}^+ f_{21}^+(r) +  c_{21}^- f_{21}^-(r) +
 c_{22}^+ f_{22}^+(r) + c_{22}^- f_{22}^-(r), \nonumber \\
 g_2(r) = \left(\frac{a }{2 \sqrt{b}}\right) (c_{21}^+ g_{21}^+(r) +  c_{21}^- g_{21}^-(r) +
 c_{22}^+ g_{22}^+(r) + c_{22}^- g_{22}^-(r)) ,
\end{eqnarray}
where
\begin{equation}
 f_{21}^{\pm}(r) = M(m+1- k^{\pm}_i,m+1,b r^2),
\quad g_{21}^{\pm}(r) = \frac{ k^{\pm}_i}{(m+1)} \frac{M(m+1-
k^{\pm}_i,m+2,b r^2)} {(-k^{\pm}_i + (4 b)^{-1} e +s - 1/2)} ,
\end{equation}
\begin{equation}
f_{22}^{\pm}(r) =  U(m+1-k_i^{\pm},m+1,b r^2),\quad
g_{22}^{\pm}(r) = \frac{U(m +1- k_i^{\pm},m+2,b r^2)}{(-k_i^{\pm}
+ (4 b)^{-1} e  +s - 1/2)}
\end{equation}
for $m=0,1,2...$ and
\begin{equation}
 f_{21}^{\pm}(r) = M(1- k^{\pm}_i,-m+1,b r^2),
\quad
 g_{21}^{\pm}(r) = m \frac{M(- k^{\pm}_i,-m,b
r^2)}{(-k^{\pm}_i + (4 b)^{-1} e +s - 1/2)} ,
\end{equation}
\begin{equation}
 f_{22}^{\pm}(r) =  U(1-k_i^{\pm},-m+1,b r^2) , \quad
g_{22}^{\pm}(r) = \frac{U(- k_i^{\pm},-m,b r^2)}{(-k_i^{\pm} + (4
b)^{-1} e  +s - 1/2)}
\end{equation}
for  $m=-1,-2,-3...$  Here the notation
\begin{equation}
k^{\pm}_i = \frac{1}{4 b} \left( e + \frac{a^2}{2} \pm a \sqrt{e
+\frac{a^2}{4}
 + \left(\frac{4 b}{a}\right)^2 (s - 1/2)^2} \right)
 \end{equation}
is introduced.

The continuity conditions for the radial wave functions $u(r),
w(r)$ and their derivatives $u'(r), w'(r)$ at the boundary points
$r=r_i$  and $r=1$ can be written in the following form
 \begin{eqnarray}
f_1(r_i) &=& f_2(r_i), \quad
  {f_1}'(r_i)= {f_2}'(r_i), \quad
 g_1(r_i) = g_2(r_i), \quad
  {g_1}'(r_i)= {g_2}'(r_i),\nonumber \\
f_2(1) &=& f_3(1), \quad
  {f_2}'(1)= {f_3}'(1), \quad
 g_2(1) = g_3(1), \quad
  {g_2}'(1)= {g_3}'(1) .
\end{eqnarray}
 Hence,  we obtain the algebraic equations
\begin{equation}
M(m,e,v,a,b,s,r_i){\bf X} = 0
\end{equation}
for eight coefficients, where   $ {\bf X} =(c_1^+, c_1^-,
c_{21}^+, c_{21}^-, c_{22}^+, c_{22}^-, c_3^+, c_3^-)  $ and
$M(m,e,v,a,b,s,r_i)$ is $8 \times 8$ matrix
\begin{equation}
 M=
  \left(\begin{array}{rrrrrrrr}
  f_1^+(r_i)&f_1^-(r_i)& -f_{21}^+(r_i)&-f_{21}^-(r_i)& -f_{22}^+(r_i)&-f_{22}^-(r_i)&
   0&0 \\
 {f_{1}^+}'(r_i)&{f_1^-}'(r_i)&
   -{f_{21}^+}'(r_i)&-{f_{21}^-}'(r_i)& -{f_{22}^+}'(r_i)&-{f_{22}^-}'(r_i)&
   0&0 \\
  g_1^+(r_i)&g_1^-(r_i)&
   -g_{21}^+(r_i)&-g_{21}^-(r_i) & -g_{22}^+(r_i)&-g_{22}^-(r_i)&
   0&0 \\
 {g_1^+}'(r_i)&{g_1^-}'(r_i)&
 -{g_{21}^+}'(r_i)&-{g_{21}^-}'(r_i)& -{g_{22}^+}'(r_i)&-{g_{22}^-}'(r_i)&
   0&0 \\
 0&0&
   f_{21}^+(1)&f_{21}^-(1) &  f_{22}^+(1)&f_{22}^-(1)&
 -f_{3}^+(1)&-f_{3}^-(1) \\
 0&0&
   {f_{21}^+}'(1)&{f_{21}^-}'(1)&  {f_{22}^+}'(1)&{f_{22}^-}'(1)&
   -{f_{3}^+}'(1)&-{f_{3}^-}'(1) \\
0&0&
   g_{21}^+(1)&g_{21}^-(1) &  g_{22}^+(1)&g_{22}^-(1)&
   -g_{3}^+(1)&-g_{3}^-(1) \\
 0&0&
   {g_{21}^+}'(1)&{g_{21}^-}'(1) &{g_{22}^+}'(1)&{g_{22}^-}'(1)&
   -{g_{3}^+}'(1)&-{g_{3}^-}'(1)
 \end{array}\right) .
\end{equation}

Thus,  the exact equation for the energy  $e(m,v,a,b,s,r_i)$ reads
 \begin{equation}
\det M(m,e,v,a,b,s,r_i) =0.
\end{equation}
This equation is solved numerically. If the energy values
$e(m,v,a,b,s,r_i)$ are found from Eq. (28), then it is simply to
get coefficients $ c_1^+, c_1^-, c_{21}^+, c_{21}^-, c_{22}^+,
c_{22}^-, c_3^+$ and $c_3^- $ from Eq. (26) and the normalization
condition $\int_0^{\infty} (u^2(r) + w^2(r))r \,dr = 1$ in order
to construct the radial wave functions completely.

\section{Numerical and graphic illustrations}

Now we present the results of of the numerical solution of Eq.
(28). In accordance with \cite{tsi} we choose the following values
of parameters $M_{eff}/M_e =0.067, g= -0.44$, then we have
$s=-0.00737$. If we assume $\rho_o = 30 \, nm$, then the
correspondences $a=1 \to a_R = 18.958 \, meV \, nm$, $b=1 \to
B=1.45649 \, T$ and $e=1 \to E= 0.63193 \, meV$ between
dimensionless and dimensional quantities are obtained. The
dimensionless depth  $v= 400$ of the potential well corresponds to
the dimensional depth $V= 252.77 \, meV$ that is close to value $V
= 257 \, meV$ in \cite{gro}. Figures correspond to ratio $r_i =
0.5$. The solid lines represent the first energy levels and the
dashed lines  represent the second energy levels for the given
values of $m$. Figures 1 and 2 show the dependence of energy
levels $e$ on magnetic field $b$ at the fixed values of the Rashba
parameter $a$. Figures 3 and 4 demonstrate the dependence of
energy levels $e$ on the Rashba parameter $a$ at the fixed values
of magnetic field $b$. Note that the character of both
dependencies is conserved if the values of fixed parameters are
varied  in wide range.

In addition, Tables represent the essential dependence of the
energy levels on the shape of ring and the depth of potential
well. Table 1 shows the dependence  of $e$ on ratio $r_i$ for the
fixed values $v=400, \,a=1, \, b=1$. Table 2 demonstrates the
dependence of $e$ on  $v$ for the fixed values $r_i=0.5, \,a=1, \,
b=1$. The first and the second columns for the given values of $m$
represent the first and the second energy levels, respectively. We
see that the energy levels increase with the growth of $r_i$ and
$v$.

\begin{figure}[p]
\centering
 \includegraphics[width=15cm]{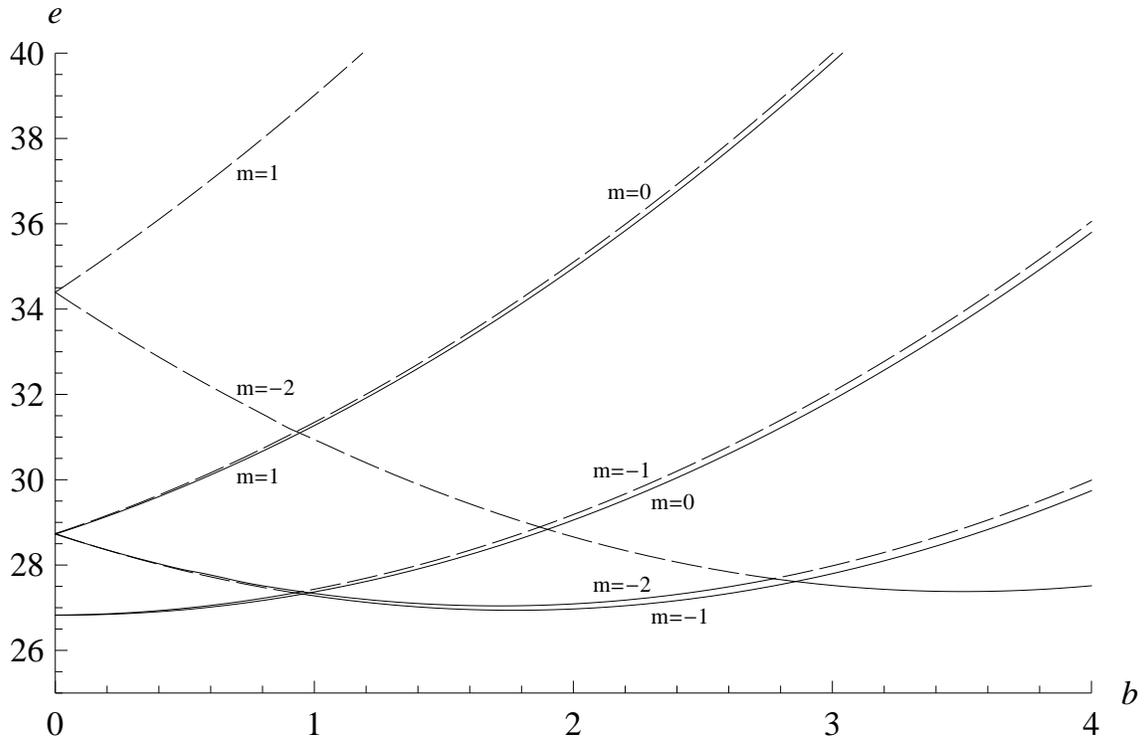} \caption{Dependence of $e$
on $b$ at $a = 0.1$.}
\end{figure}
\begin{figure}[p]
\centering
  \includegraphics[width=15cm]{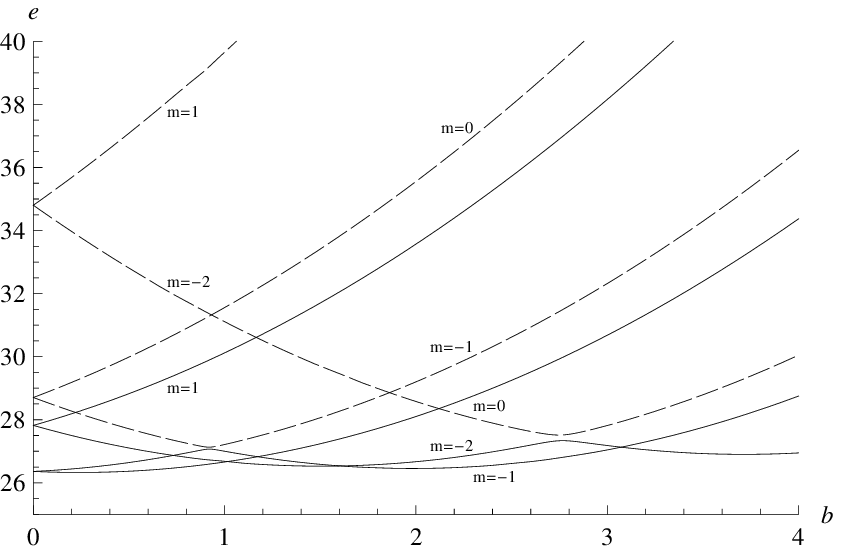} \caption{Dependence of $e$
on $b$ at $a = 1$.}
\end{figure}

\begin{figure}[p]
\centering
 \includegraphics[width=15cm]{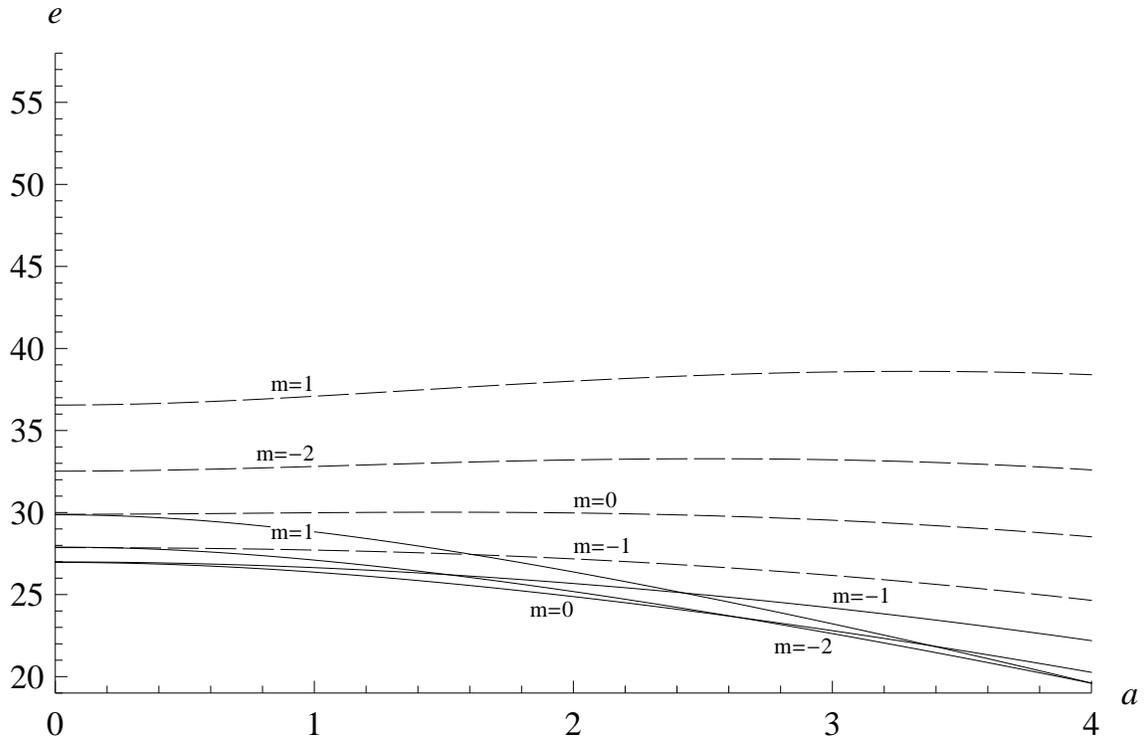} \caption{Dependence of $e$
on $a$ at $b = 0.5$.}
\end{figure}
\begin{figure}[p]
\centering
  \includegraphics[width=15cm]{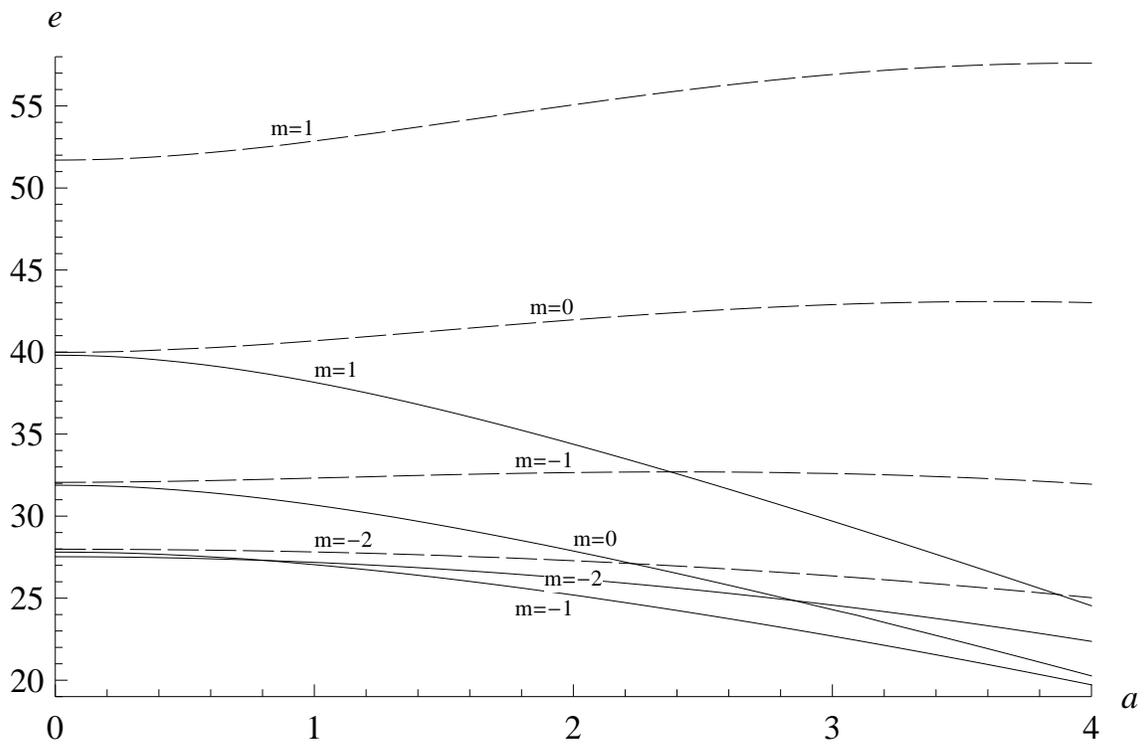} \caption{Dependence of $e$
on $a$ at $b = 3$.}
\end{figure}

\newpage
\section{Conclusion}

In our opinion, the examined  exactly solvable  model  with the
realistic potential well of finite depth is physically adequate in
order to describe the behavior  of an electron in a semiconductor
quantum ring of finite width with account of the Rashba spin-orbit
interaction in the presence of an external magnetic field.

\begin{table}[t]
\caption{Dependence of $e$ on $r_i$  at $v=400, \, a=1, \, b=1$.}
\begin{center}
\begin{tabular}{r|*8r}
 \hline
$r_{i}$ & \multicolumn{2}{c}{$m=-2$} & \multicolumn{2}{c}{$m=-1$} & \multicolumn{2}{c}{$m=0$} & \multicolumn{2}{c}{$m=1$}  \\
 \cline{2-9}
  & \multicolumn{8}{c}{$e$}\\
  \hline
 0.1 & 11.1546 & 20.5634 & 8.40784 & 11.6889 & 8.07216 & 16.0687 & 14.7685 & 29.0068 \\
 0.3 & 15.2333 & 22.0607 & 14.8695 & 15.6909 & 14.4308 & 20.1349 & 18.7835 & 30.5326 \\
 0.5 & 26.6594 & 31.1076 & 27.0008 & 27.2145 & 26.6594 & 31.5576 & 30.1207 & 39.6641 \\
 0.7 & 60.0830 & 62.9589 & 60.2745 & 61.0963 & 60.4248 & 64.9463 & 63.4123 & 71.6304 \\
 0.9 & 217.766 & 219.587 & 217.804 & 219.066 & 218.264 & 222.606 & 220.965 & 228.388 \\
 \hline
\end{tabular}
\end{center}
\end{table}

\begin{table}[t]
\caption{ Dependence of $e$ on $v$ at $r_{i}=0.5, \, a=1, \,
b=1$.}
\begin{center}
\begin{tabular}{r|*8r}
 \hline
$v$ & \multicolumn{2}{c}{$m=-2$} & \multicolumn{2}{c}{$m=-1$} & \multicolumn{2}{c}{$m=0$} & \multicolumn{2}{c}{$m=1$}  \\
 \cline{2-9}
  & \multicolumn{8}{c}{$e$}\\
  \hline
 25   & 11.1356 & 15.5883 & 11.0065 & 11.6889 & 10.4833 & 15.9651 & 14.5421 & 20.6312 \\
 50   & 15.0121 & 19.6818 & 15.2366 & 15.3623 & 14.7204 & 19.8905 & 18.4485 & 28.2691 \\
 100  & 19.1740 & 23.7577 & 19.4880 & 19.6275 & 19.0631 & 24.0567 & 22.6185 & 32.3247 \\
 400  & 26.6724 & 31.1076 & 27.0008 & 27.2145 & 26.6594 & 31.5576 & 30.1207 & 39.6634 \\
 1000 & 30.4209 & 34.8152 & 30.7516 & 30.9818 & 30.4279 & 35.3066 & 33.8699 & 43.3697 \\
\hline
\end{tabular}
\end{center}
\end{table}


\begin{thebibliography}{xxxx}
\bibitem{ban} T.V. Bandos, A. Cantarero and A. Garcia-Crist\'obal, Eur. Phys.
J. B \textbf{53}, 99-108 (2006).
\bibitem{gro} M. Grochol, F. Grosse and R. Zimmermann, Phys. Rev. B \textbf{74}, 115416 (2006).
\bibitem{chap} A.\,V. Chaplik and L.\,I. Magarill, Phys. Rev. Lett. \textbf{96}, 126402 (2006).
\bibitem{kud1} V.V. Kudryashov,  Proc. of the XIII Intern. School-Conference "Foundations and
 Advances in Nonlinear Science", Minsk, 2006, pp. 125-130; Nonlinear Phenomena in Complex
Systems \textbf{12}, 199-203 (2009).
\bibitem{ras} E.I. Rashba,
Fiz. Tverd. Tela (Leningrad) \textbf{2}, 1224 (1960) [Sov. Phys.
Solid State \textbf{2}, 1109 (1960)].
\bibitem{byc} Yu.A. Bychkov and E.I. Rashba,  J. Phys. C \textbf{17},
6039-6046 (1984).
\bibitem{kud2} V.V. Kudryashov,  Proc. of the XIV Intern. School-Conference "Foundations and
 Advances in Nonlinear Science", Minsk, 2009, pp. 95-102.
\bibitem{bul} E.N. Bulgakov and A.F. Sadreev,  Pis'ma v ZhETF \textbf{73},
573 2001) [JETP Lett. \textbf{73}, 505 (2001)].
\bibitem{tsi} E. Tsitsishvili, G.S. Lozano and A.O. Gogolin,
Phys. Rev. B \textbf{70}, 115316 (2004).
\bibitem{abr}  M. Abramovitz and I.A. Stegun (eds.),
  \textit{Handbook of Mathematical Functions} ( Dover Publications, New
 York, 1970).
\end{thebibliography}
\end{document}